\documentclass[10pt]{article}
\usepackage{amsmath, cancel, fullpage, mathtools, mdframed, subfig, txfonts, blkarray, multirow}
\mdfdefinestyle{style1}{skipabove=0.5cm, skipbelow=0.5cm, leftmargin=0cm, rightmargin=0cm, linewidth=1pt}

\newcommand{\sgn}{\operatorname{sgn}}

\begin{document}
    \title{Space-Time Finite Element for Sensor Fusion}

    \author{Markus Pagitz\\
            markus@pagitz.de}
    \date{}
    \maketitle

    \begin{abstract}
        Drones estimate their position and orientation with the help of various sensors. Their data streams, that differ with respect to the sampling rate and standard deviation, need to be fused to get an accurate position and orientation estimate. It is subsequently shown that a nonlinear space-time finite element and static condensation can be used to accomplish this task. This is done, for the sake of clarity, in three stages. The first stage estimates the local magnetic north vector with the help of magnetometers and gyroscopes. The second stage projects the remaining sensor data onto the plane that is orthogonal to the local magnetic north vector and the third stage solves the corresponding two-dimensional problem.
    \end{abstract}


    \section{Introduction}
        Drones are fascinating machines. Various kind of relatively inexpensive sensors and the corresponding fusion algorithms are the main components that enable their remarkable flight performances. This manuscript originated from a structural engineers pastime interest in this field. The subsequently presented approach is solely based on methods that are widely used in the development of finite elements. It might therefore be of help for other people that do not possess a comprehensive background in nonlinear fusion algorithms such as the Extended Kalman Filter \cite{McElhoe1966, Smith1962}, Unscented Kalman Filter \cite{Julier1997} and Particle Methods \cite{DelMoral1996}.\\

        The position of a drone can be reliably estimated with the help of a GNSS receiver whose estimates are improved with the help of various sensors such as a barometer, magnetometers, accelerometers and gyroscopes. This is subsequently done in the following way. The magnetic north vector in a drone fixed coordinate system can be estimated with the help of magnetometers and gyroscopes. This vector can then be used to span an orthogonal plane onto which the remaining sensor data is projected. The rotation around the magnetic north vector and an improved position estimate can then be computed from the resulting two-dimensional subproblem.\footnote{It should be noted that the position and orientation of a drone can be computed in three stages or, as a fully coupled problem, in a single step. The former approach has the advantage that it leads to a clearer presentation so that it is subsequently used.}


    \section{Magnetic North}

        \subsection{Sensor Data}
            The local magnetic north vector can be estimated with the help of magnetometers and gyroscopes. The data stream of the three-axis magnetometer for a single finite element in the drone fixed coordinate system is
            \begin{align}
                \left[
                \begin{array}{c}
                    \mathbf{t}_m\\
                    \mathbf{x}_m\\
                    \mathbf{y}_m\\
                    \mathbf{z}_m
                \end{array}
                \right] &=
                \left[
                \begin{array}{ccccc}
                    t_{m,1} & t_{m,2} & t_{m,3} & \cdots & t_{m,n_m}\\
                    x_{m,1} & x_{m,2} & x_{m,3} & \cdots & x_{m,n_m}\\
                    y_{m,1} & y_{m,2} & y_{m,3} & \cdots & y_{m,n_m}\\
                    z_{m,1} & z_{m,2} & z_{m,3} & \cdots & z_{m,n_m}
                \end{array}
                \right]
            \end{align}
            where the sampling rate is usually about 100~Hz. It should be noted that the mean vector length ${x_{m,i}}^2+{y_{m,i}}^2+{z_{m,i}}^2$ is assumed to possess a unit length. The corresponding data stream of the gyroscopes is
            \begin{align}
                \left[
                \begin{array}{c}
                    \mathbf{t}_g\\
                    \dot{\boldsymbol{\varphi}}_{x,g}\\
                    \dot{\boldsymbol{\varphi}}_{y,g}\\
                    \dot{\boldsymbol{\varphi}}_{z,g}
                \end{array}
                \right] &=
                \left[
                \begin{array}{ccccc}
                    t_{g,1} & t_{g,2} & t_{g,3} & \cdots & t_{g,n_g}\\
                    \dot{\varphi}_{x,g,1} & \dot{\varphi}_{x,g,2} & \dot{\varphi}_{x,g,3} & \cdots & \dot{\varphi}_{x,g,n_g}\\
                    \dot{\varphi}_{y,g,1} & \dot{\varphi}_{y,g,2} & \dot{\varphi}_{y,g,3} & \cdots & \dot{\varphi}_{y,g,n_g}\\
                    \dot{\varphi}_{z,g,1} & \dot{\varphi}_{z,g,2} & \dot{\varphi}_{z,g,3} & \cdots & \dot{\varphi}_{z,g,n_g}
                \end{array}
                \right]
            \end{align}
            where the sampling rate is usually about 1~kHz. Interpolation of the sensor data in the domain of a finite element is done with the help of Hermitian polynomials\footnote{See Appendix for details.} $\mathbf{N}$ so that, for example
            \begin{align}
                \mathbf{x}^\prime_m = \mathbf{N}_m \mathbf{R}_x \mathbf{u} = \mathbf{N}_{m,x} \mathbf{u}
            \end{align}
            where $\mathbf{x}^\prime_m$ is the interpolated magnetometer data, $\mathbf{R}_x$ selects the relevant degrees of freedom from $\mathbf{u}$ and $\mathbf{N}_m = \mathbf{N}\left(\mathbf{t}_m\right)$.


        \subsection{Error Terms}
            The interpolation error $E_S$ of the sensor data can be written as
            \begin{align}
                E_s =&\ \ \ \ \frac{1}{2 \sigma_m^2} \left(\left(\mathbf{x}^\prime_m - \mathbf{x}_m\right)^\top \left(\mathbf{x}^\prime_m - \mathbf{x}_m\right) + \left(\mathbf{y}^\prime_m - \mathbf{y}_m\right)^\top \left(\mathbf{y}^\prime_m - \mathbf{y}_m\right) + \left(\mathbf{z}^\prime_m - \mathbf{z}_m\right)^\top \left(\mathbf{z}^\prime_m - \mathbf{z}_m\right)\right)\\\nonumber
                & + \frac{1}{2 \overline{\sigma}_g^2} \left( \left(\dot{\mathbf{x}}^\prime_g - \dot{\mathbf{x}}_g\right)^\top \left(\dot{\mathbf{x}}^\prime_g - \dot{\mathbf{x}}_g\right)
                + \left(\dot{\mathbf{y}}^\prime_g - \dot{\mathbf{y}}_g\right)^\top \left(\dot{\mathbf{y}}^\prime_g - \dot{\mathbf{y}}_g\right)
                + \left(\dot{\mathbf{z}}^\prime_g - \dot{\mathbf{z}}_g\right)^\top \left(\dot{\mathbf{z}}^\prime_g - \dot{\mathbf{z}}_g\right) \right)
            \end{align}
            where
            \begin{align}
                \left[
                \begin{array}{c}
                    \dot{x}_{g,i}\\
                    \dot{y}_{g,i}\\
                    \dot{z}_{g,i}
                \end{array}
                \right] =
                \left[
                \begin{array}{c}
                    x_{g,i}^\prime\\
                    y_{g,i}^\prime\\
                    z_{g,i}^\prime
                \end{array}
                \right]
                \times
                \left[
                \begin{array}{c}
                    \dot{\varphi}_{x,g,i}\\
                    \dot{\varphi}_{y,g,i}\\
                    \dot{\varphi}_{z,g,i}
                \end{array}
                \right] =
                \left[
                \begin{array}{c}
                    y_{g,i}^\prime \dot{\varphi}_{z,g,i} - z_{g,i}^\prime \dot{\varphi}_{y,g,i}\\
                    z_{g,i}^\prime \dot{\varphi}_{x,g,i} - x_{g,i}^\prime \dot{\varphi}_{z,g,i}\\
                    x_{g,i}^\prime \dot{\varphi}_{y,g,i} - y_{g,i}^\prime \dot{\varphi}_{x,g,i}
                \end{array}
                \right].
            \end{align}
            The corresponding first and second derivatives of the error term are
            \begin{align}
                \frac{\partial E_s}{\partial \mathbf{u}} = &\hspace{3.5mm} \frac{1}{\sigma_m^2} \left(
                {\mathbf{N}_{m,x}}^\top
                \left(\mathbf{N}_{m,x} \mathbf{u} - \mathbf{x}_m\right) +
                {\mathbf{N}_{m,y}}^\top
                \left(\mathbf{N}_{m,y} \mathbf{u} - \mathbf{y}_m\right) +
                {\mathbf{N}_{m,z}}^\top
                \left(\mathbf{N}_{m,z} \mathbf{u} - \mathbf{z}_m\right)\right)\\\nonumber
                &+ \frac{1}{\overline{\sigma}_g^2} \left(
                {\dot{\mathbf{N}}_{g,x}}^\top
                \left(\dot{\mathbf{N}}_{g,x} \mathbf{u} - \dot{\mathbf{x}}_g\right) +
                {\dot{\mathbf{N}}_{g,y}}^\top
                \left(\dot{\mathbf{N}}_{g,y} \mathbf{u} - \dot{\mathbf{y}}_g\right) +
                {\dot{\mathbf{N}}_{g,z}}^\top
                \left(\dot{\mathbf{N}}_{g,z} \mathbf{u} - \dot{\mathbf{z}}_g\right)\right)
            \end{align}
            and
            \begin{align}
                \frac{\partial^2 E_s}{\partial \mathbf{u}^2} = &\ \ \ \frac{1}{\sigma_m^2} \left(
                {\mathbf{N}_{m,x}}^\top \mathbf{N}_{m,x} +
                {\mathbf{N}_{m,y}}^\top \mathbf{N}_{m,y} +
                {\mathbf{N}_{m,z}}^\top \mathbf{N}_{m,z} \right) +
                \frac{1}{\sigma_g^2}\left( {\dot{\mathbf{N}}_{g,x}}^\top
                \dot{\mathbf{N}}_{g,x} + {\dot{\mathbf{N}}_{g,y}}^\top
                \dot{\mathbf{N}}_{g,y} + {\dot{\mathbf{N}}_{g,z}}^\top
                \dot{\mathbf{N}}_{g,z} \right).
            \end{align}
            It should be noted that a single variance for the second error term is used to simplify the presentation. In reality, the terms $\dot{x}_g$, $\dot{y}_g$, $\dot{z}_g$ possess an independent, non-constant variance that needs to be computed from $\sigma_g$, $\sigma_m$ and the current state $x^\prime_g$, $y^\prime_g$, $z^\prime_g$ as well as $\dot{\varphi}_{x,g}$, $\dot{\varphi}_{y,g}$, $\dot{\varphi}_{z,g}$. However, it will later be shown that this simplification has only a small impact on the orientation and position estimates.


        \subsection{Length Constraint}
            Minimization of the error terms leads to expressions where the mean vector length ${x_i}^2 + {y_i}^2 + {z_i}^2$ is smaller than one. This undesired effect is driven by the dominant, gyroscope related terms that become smaller for decreasing vector lengths. It is therefore necessary to enforce a mean unit length with the help of an additional constraint. The interpolated vector lengths $\mathbf{L}$ at the gyroscope readings are
            \begin{align}
                \mathbf{L} = \left(\mathbf{x}_g^\prime \odot \mathbf{x}_g^\prime + \mathbf{y}_g^\prime \odot \mathbf{y}_g^\prime + \mathbf{z}_g^\prime \odot \mathbf{z}_g^\prime\right)^{\odot \frac{1}{2}}
            \end{align}
            so that the corresponding first and second derivatives are
            \begin{align}
                \frac{\partial \mathbf{L}}{\partial \mathbf{u}} = \textrm{diag}\left(\mathbf{N}_{g,x} \mathbf{u} \odot \mathbf{L}^{\odot\left(-1\right)}\right) \mathbf{N}_{g,x} + \textrm{diag}\left(\mathbf{N}_{g,y} \mathbf{u} \odot \mathbf{L}^{\odot\left(-1\right)}\right) \mathbf{N}_{g,y} + \textrm{diag}\left(\mathbf{N}_{g,z} \mathbf{u} \odot \mathbf{L}^{\odot\left(-1\right)}\right) \mathbf{N}_{g,z}
            \end{align}
            and
            \begin{align}
                \frac{\partial^2 L_i}{\partial \mathbf{u}^2} = \frac{1}{L_i} \left({\mathbf{N}_{g,x,i}}^\top \mathbf{N}_{g,x,i} + {\mathbf{N}_{g,y,i}}^\top \mathbf{N}_{g,y,i} + {\mathbf{N}_{g,z,i}}^\top \mathbf{N}_{g,z,i} - \frac{\partial L_i}{\partial \mathbf{u}}^\top \frac{\partial L_i}{\partial \mathbf{u}}\right).
            \end{align}
            This enables the definition of an additional error term
            \begin{align}
                E_L = \frac{1}{2 \sigma_L^2} \left(\mathbf{L}-\mathbf{1}\right)^\top \left(\mathbf{L}-\mathbf{1}\right)
            \end{align}
            whose derivatives are
            \begin{align}
                \frac{\partial E_L}{\partial \mathbf{u}} = \frac{1}{\sigma_L^2} \frac{\partial \mathbf{L}}{\partial \mathbf{u}}^\top \left(\mathbf{L} - \mathbf{1}\right)
            \end{align}
            and
            \begin{align}
                \frac{\partial^2 E_L}{\partial \mathbf{u}^2} &= \frac{1}{\sigma_L^2} \left( \frac{\partial \mathbf{L}}{\partial \mathbf{u}}^\top \frac{\partial \mathbf{L}}{\partial \mathbf{u}} + \sum_{i=1}^{n_g} \frac{\partial^2 L_i}{\partial \mathbf{u}^2} \left(L_i - 1\right)\right)\\\nonumber
                &=
                \frac{1}{\sigma_L^2} \left(
                {\mathbf{N}_{g,x,i}}^\top \textrm{diag}\left(\mathbf{1}-\mathbf{L}^{\odot\left(-1\right)}\right)
                \mathbf{N}_{g,x,i} + \ldots +
                \frac{\partial \mathbf{L}}{\partial \mathbf{u}}^\top
                \textrm{diag}\left(\mathbf{L}^{\odot\left(-1\right)}\right)
                \frac{\partial \mathbf{L}}{\partial \mathbf{u}}\right)
            \end{align}
            since
            \begin{align}
                \sum_{i=1}^{n_g} \frac{\partial^2 L_i}{\partial \mathbf{u}^2} \left(L_i-1\right) = {\mathbf{N}_{g,x,i}}^\top \textrm{diag}\left(\mathbf{1}-\mathbf{L}^{\odot\left(-1\right)}\right) \mathbf{N}_{g,x,i} + \ldots - \frac{\partial \mathbf{L}}{\partial \mathbf{u}}^\top  \textrm{diag}\left(\mathbf{1}-\mathbf{L}^{\odot\left(-1\right)}\right) \frac{\partial \mathbf{L}}{\partial \mathbf{u}}.
            \end{align}
            Based on numerical considerations, the standard deviation $\sigma_L$ should be chosen as large as possible. Instead of using a penalty method it is possible to exactly enforce the error term $\partial E_L/\partial \mathbf{u} = \mathbf{0}$ with the help of Lagrange multipliers. However, this increases the total number of degrees of freedom and slightly complicates the subsequently presented static condensation. Furthermore, the length constraint can not be exactly enforced due to the finite number of shape functions so that an additional normalization of the interpolated magnetic north vector is required anyhow.


        \subsection{Solution Approach}
            The total error of the $i$-th finite element is composed of the sensor data and the length constraint so that
            \begin{align}
                E^i = E_s^i + E_L^i.
            \end{align}
            The degrees of freedom $\mathbf{u}$ can be split into $\mathbf{u}_1$, $\mathbf{u}_2$ that are located at the left, right element boundary respectively. Hence, the augmented derivatives of the error terms can be written as
            \begin{align}
                \frac{\partial E^i}{\partial \mathbf{u}} =
                \left[
                \begin{array}{cc}
                    \displaystyle \frac{\partial E^i}{\partial \mathbf{u}_1}^\top + \left(\mathbf{u}^i_1 - \mathbf{u}^{i-1}_2\right)^\top \mathbf{K}^{i-1} & \displaystyle \frac{\partial E^i}{\partial \mathbf{u}_2}^\top
                \end{array}
                \right]^\top
            \end{align}
            and
            \begin{align}
                \frac{\partial^2 E^i}{\partial \mathbf{u}^2} =
                \left[
                \begin{array}{cc}
                    \displaystyle \frac{\partial^2 E^i}{\partial {\mathbf{u}_1}^2} + \mathbf{K}^{i-1} & \displaystyle \frac{\partial^2 E^i}{\partial \mathbf{u}_1 \partial \mathbf{u}_2}\\
                    \displaystyle \frac{\partial^2 E^i}{\partial \mathbf{u}_2 \partial \mathbf{u}_1} & \displaystyle \frac{\partial^2 E^i}{\partial {\mathbf{u}_2}^2}
                \end{array}
                \right]
            \end{align}
            where the condensed, symmetric error matrix $\mathbf{K}^{i-1}$ of the $\left(i-1\right)$-th finite element with respect to the degrees of freedom $\mathbf{u}_2^{i-1}$ is
            \begin{align}
                \mathbf{K}^{i-1} = \frac{\partial^2 E^{i-1}}{\partial {\mathbf{u}_2}^2} - \frac{\partial^2 E^{i-1}}{\partial \mathbf{u}_2 \partial \mathbf{u}_1} \left(\frac{\partial^2 E^{i-1}}{\partial {\mathbf{u}_1}^2}\right)^{-1} \frac{\partial^2 E^{i-1}}{\partial \mathbf{u}_1 \partial \mathbf{u}_2}.
            \end{align}
            The condensed error matrix \cite{Guyan1965} is the gain from the interpolation of the previous data that was processed with one or more finite elements. The resulting system of nonlinear equations\footnote{The nonlinear part stems solely from the length constraint so that the nonlinearity of these equations is relatively mild.} can be solved for $\mathbf{u}^i$ with the help of a Newton based approach within a few iterations. It should be noted that the required number of iterations can be significantly reduced if the degrees of freedom from the previous element are extrapolated prior to the minimization.


    \section{Projection}
        The local magnetic north vector is known from the sensor fusion of gyroscopes and magnetometers whereas the global magnetic north vector is known from maps or measurements. It is therefore possible to project the remaining sensor data onto a plane that is orthogonal to the local magnetic north vector.


        \subsection{Local Coordinate System}
            The normalized magnetic north vector that is subsequently denoted as $\mathbf{x}^\prime$ can be written at the gyroscope readings as
            \begin{align}
                \mathbf{x}_g^\prime &=
                \left[
                \begin{array}{ccc}
                    \left(\mathbf{N}_g \mathbf{R}_x \mathbf{u}\right) \odot \mathbf{L}_{x,g}^{\odot\left(-1\right)} &
                    \left(\mathbf{N}_g \mathbf{R}_y \mathbf{u}\right) \odot \mathbf{L}_{x,g}^{\odot\left(-1\right)} &
                    \left(\mathbf{N}_g \mathbf{R}_z \mathbf{u}\right) \odot \mathbf{L}_{x,g}^{\odot\left(-1\right)}
                \end{array}
                \right]^\top\\\nonumber
                \dot{\mathbf{x}}_g^\prime &=\hspace{2.5mm}
                \left[
                \begin{array}{ccc}
                    \left(\dot{\mathbf{N}}_g \mathbf{R}_x \mathbf{u}\right) \odot \mathbf{L}_{x,g}^{\odot\left(-1\right)} &
                    \left(\dot{\mathbf{N}}_g \mathbf{R}_y \mathbf{u}\right) \odot \mathbf{L}_{x,g}^{\odot\left(-1\right)} &
                    \left(\dot{\mathbf{N}}_g \mathbf{R}_z \mathbf{u}\right) \odot \mathbf{L}_{x,g}^{\odot\left(-1\right)}
                \end{array}
                \right]^\top\\\nonumber
                &\hspace{3mm}-
                \left[
                \begin{array}{ccc}
                    \left(\mathbf{N}_g \mathbf{R}_x \mathbf{u}\right) \odot \dot{\mathbf{L}}_{x,g} \odot \mathbf{L}_{x,g}^{\odot\left(-2\right)} &
                    \left(\mathbf{N}_g \mathbf{R}_y \mathbf{u}\right) \odot \dot{\mathbf{L}}_{x,g} \odot \mathbf{L}_{x,g}^{\odot\left(-2\right)} &
                    \left(\mathbf{N}_g \mathbf{R}_z \mathbf{u}\right) \odot \dot{\mathbf{L}}_{x,g} \odot \mathbf{L}_{x,g}^{\odot\left(-2\right)}
                \end{array}
                \right]^\top
            \end{align}
            where the vector length is
            \begin{align}
                \mathbf{L}_{x,g} = \left(\left(\mathbf{N}_g \mathbf{R}_x \mathbf{u}\right)^{\odot 2} + \left(\mathbf{N}_g \mathbf{R}_y \mathbf{u}\right)^{\odot 2} + \left(\mathbf{N}_g \mathbf{R}_z \mathbf{u}\right)^{\odot 2}\right)^{\odot \frac{1}{2}}
            \end{align}
            and its time derivative is
            \begin{align}
                \dot{\mathbf{L}}_{x,g} = \left(\left(\dot{\mathbf{N}}_g \mathbf{R}_x \mathbf{u}\right) \odot \left(\mathbf{N}_g \mathbf{R}_x \mathbf{u}\right) + \left(\dot{\mathbf{N}}_g \mathbf{R}_y \mathbf{u}\right) \odot \left(\mathbf{N}_g \mathbf{R}_y \mathbf{u}\right) + \left(\dot{\mathbf{N}}_g \mathbf{R}_z \mathbf{u}\right) \odot \left(\mathbf{N}_g \mathbf{R}_z \mathbf{u}\right) \right) \odot \mathbf{L}_{x,g}^{\odot \left(-1\right)}.
            \end{align}
            The unit vector $\mathbf{y}^\prime$ that is orthogonal to $\mathbf{x}^\prime$ and the global magnetic north vector $\mathbf{d}$ is
            \begin{align}
                \mathbf{y}^\prime &=
                \left[
                \begin{array}{c}
                    \left(\mathbf{x}_2^\prime d_3 - \mathbf{x}_3^\prime d_2\right) \odot \mathbf{L}_y^{\odot \left(-1\right)}\\
                    \left(\mathbf{x}_3^\prime d_1 - \mathbf{x}_1^\prime d_3\right) \odot \mathbf{L}_y^{\odot \left(-1\right)}\\
                    \left(\mathbf{x}_1^\prime d_2 - \mathbf{x}_2^\prime d_1\right) \odot \mathbf{L}_y^{\odot \left(-1\right)}
                \end{array}
                \right]\\\nonumber
                \dot{\mathbf{y}}^\prime &=
                \left[
                \begin{array}{c}
                    \left(\dot{\mathbf{x}}^\prime_2 d_3 - \dot{\mathbf{x}}^\prime_3 d_2\right) \odot \mathbf{L}_y^{\odot \left(-1\right)}\\
                    \left(\dot{\mathbf{x}}^\prime_3 d_1 - \dot{\mathbf{x}}^\prime_1 d_3\right) \odot \mathbf{L}_y^{\odot \left(-1\right)}\\
                    \left(\dot{\mathbf{x}}^\prime_1 d_2 - \dot{\mathbf{x}}^\prime_2 d_1\right) \odot \mathbf{L}_y^{\odot \left(-1\right)}
                \end{array}
                \right] -
                \left[
                \begin{array}{c}
                    \left(\mathbf{x}_2^\prime d_3 - \mathbf{x}_3^\prime d_2\right) \odot \dot{\mathbf{L}}_y \odot \mathbf{L}_y^{\odot \left(-2\right)}\\
                    \left(\mathbf{x}_3^\prime d_1 - \mathbf{x}_1^\prime d_3\right) \odot \dot{\mathbf{L}}_y \odot \mathbf{L}_y^{\odot \left(-2\right)}\\
                    \left(\mathbf{x}_1^\prime d_2 - \mathbf{x}_2^\prime d_1\right) \odot \dot{\mathbf{L}}_y \odot \mathbf{L}_y^{\odot \left(-2\right)}
                \end{array}
                \right]
            \end{align}
            where
            \begin{align}
                \mathbf{L}_y &= \left(\left(\mathbf{x}_2^\prime d_3 - \mathbf{x}_3^\prime d_2\right)^{\odot 2} + \left(\mathbf{x}_3^\prime d_1 - \mathbf{x}_1^\prime d_3\right)^{\odot 2} + \left(\mathbf{x}_1^\prime d_2 - \mathbf{x}_2^\prime d_1\right)^{\odot 2}\right)^{\odot\frac{1}{2}}\\\nonumber
                \dot{\mathbf{L}}_y &= \left(\left(\dot{\mathbf{x}}_2^\prime d_3 - \dot{\mathbf{x}}_3^\prime d_2\right) \odot \left(\mathbf{x}_2^\prime d_3 - \mathbf{x}_3^\prime d_2\right) + \left(\dot{\mathbf{x}}_3^\prime d_1 - \dot{\mathbf{x}}_1^\prime d_3\right) \odot \left(\mathbf{x}_3^\prime d_1 - \mathbf{x}_1^\prime d_3\right) + \left(\dot{\mathbf{x}}_1^\prime d_2 - \dot{\mathbf{x}}_2^\prime d_1\right) \odot \left(\mathbf{x}_1^\prime d_2 - \mathbf{x}_2^\prime d_1\right)\right) \odot \mathbf{L}_y^{\odot \left(-1\right)}.
            \end{align}
            The vector $\mathbf{y}^\prime$ is based on a cross product and therefore might change direction between consecutive sensor readings. In order to avoid any discontinuities it is necessary that
            \begin{align}
                \mathbf{y}^{\prime\ i} = \sgn\left({{\mathbf{y}^{\prime\ i-1}}^\top \mathbf{y}^{\prime\ i}}\right) \mathbf{y}^{\prime\ i}
            \end{align}
            holds across element boundaries and within an element for all sensor readings $i \in \left[2,n\right]$. Finally, the unit vector $\mathbf{z}^\prime$ that is orthogonal to $\mathbf{x}^\prime$ and $\mathbf{y}^\prime$ is
            \begin{align}
                \mathbf{z}^\prime =
                \left[
                \begin{array}{c}
                    \mathbf{x}_2^\prime \odot \mathbf{y}_3^\prime - \mathbf{x}_3^\prime \odot \mathbf{y}_2^\prime\\
                    \mathbf{x}_3^\prime \odot \mathbf{y}_1^\prime - \mathbf{x}_1^\prime \odot \mathbf{y}_3^\prime\\
                    \mathbf{x}_1^\prime \odot \mathbf{y}_2^\prime - \mathbf{x}_2^\prime \odot \mathbf{y}_1^\prime
                \end{array}
                \right].
            \end{align}


        \subsection{Global Coordinate System}
            In a similar manner it is possible to define a global coordinate system that is aligned to the global magnetic north vector $\mathbf{d}$ so that
            \begin{align}
                \mathbf{X}^\prime = \mathbf{d} \mathbf{1}^\top
            \end{align}
            where $\mathbf{1}$ is a vector of ones with a length equal to $\mathbf{x}$, $\mathbf{y}$ and $\mathbf{z}$. The global $\mathbf{Y}$ axis is defined as
            \begin{align}
                \mathbf{Y}^\prime = \mathbf{y}^\prime
            \end{align}
            and the global $\mathbf{Z}^\prime$ axis results in
            \begin{align}
                \mathbf{Z}^\prime =
                \left[
                \begin{array}{c}
                    \mathbf{X}_2^\prime \odot \mathbf{Y}_3^\prime - \mathbf{X}_3^\prime \odot \mathbf{Y}_2^\prime\\
                    \mathbf{X}_3^\prime \odot \mathbf{Y}_1^\prime - \mathbf{X}_1^\prime \odot \mathbf{Y}_3^\prime\\
                    \mathbf{X}_1^\prime \odot \mathbf{Y}_2^\prime - \mathbf{X}_2^\prime \odot \mathbf{Y}_1^\prime
                \end{array}
                \right].
            \end{align}


        \subsection{Transformation}
            The projection of the sensor data onto the plane and the computation of the angular velocity around the local magnetic north vector is subsequently presented. The accelerations in the local coordinate system that is aligned to the local magnetic north vector are
            \begin{align}
                \left[
                \begin{array}{c}
                    \ddot{\mathbf{x}}_a^\prime\\
                    \ddot{\mathbf{y}}_a^\prime\\
                    \ddot{\mathbf{z}}_a^\prime
                \end{array}
                \right] =
                \left[
                \begin{array}{c}
                    \mathbf{x}_{1,a}^\prime \odot \ddot{\mathbf{x}}_a + \mathbf{x}_{2,a}^\prime \odot \ddot{\mathbf{y}}_a + \mathbf{x}_{3,a}^\prime \odot \ddot{\mathbf{z}}_a\\
                    \mathbf{y}_{1,a}^\prime \odot \ddot{\mathbf{x}}_a + \mathbf{y}_{2,a}^\prime \odot \ddot{\mathbf{y}}_a + \mathbf{y}_{3,a}^\prime \odot \ddot{\mathbf{z}}_a\\
                    \mathbf{z}_{1,a}^\prime \odot \ddot{\mathbf{x}}_a + \mathbf{z}_{2,a}^\prime \odot \ddot{\mathbf{y}}_a + \mathbf{z}_{3,a}^\prime \odot \ddot{\mathbf{z}}_a
                \end{array}
                \right].
            \end{align}
            The angular velocity around the local, magnetic north vector is
            \begin{align}
                \dot{\boldsymbol{\varphi}}_g^\prime = &-\mathbf{x}_{1,g}^\prime \odot \left(\boldsymbol{\dot{\varphi}_{x,g}} + 2 \sin\left(\frac{1}{2} \boldsymbol{\alpha}_g\right) \odot \left(\cos\left(\frac{1}{2} \boldsymbol{\alpha}_g\right) \odot \dot{\mathbf{y}}_{1,g}^\prime + \sin\left(\frac{1}{2} \boldsymbol{\alpha}_g\right) \odot \left(\mathbf{y}_{2,g}^\prime \odot \dot{\mathbf{y}}_{3,g}^\prime - \mathbf{y}_{3,g}^\prime \odot \dot{\mathbf{y}}_{2,g}^\prime\right) \right)\right)\\\nonumber
                &-\mathbf{x}_{2,g}^\prime \odot \left(\boldsymbol{\dot{\varphi}_{y,g}} + 2 \sin\left(\frac{1}{2} \boldsymbol{\alpha}_g\right) \odot \left(\cos\left(\frac{1}{2} \boldsymbol{\alpha}_g\right) \odot \dot{\mathbf{y}}_{2,g}^\prime + \sin\left(\frac{1}{2} \boldsymbol{\alpha}_g\right) \odot \left(\mathbf{y}_{3,g}^\prime \odot \dot{\mathbf{y}}_{1,g}^\prime - \mathbf{y}_{1,g}^\prime \odot \dot{\mathbf{y}}_{3,g}^\prime\right) \right)\right)\\\nonumber
                &-\mathbf{x}_{3,g}^\prime \odot \left(\boldsymbol{\dot{\varphi}_{z,g}} + 2 \sin\left(\frac{1}{2} \boldsymbol{\alpha}_g\right) \odot \left(\cos\left(\frac{1}{2} \boldsymbol{\alpha}_g\right) \odot \dot{\mathbf{y}}_{3,g}^\prime + \sin\left(\frac{1}{2} \boldsymbol{\alpha}_g\right) \odot \left(\mathbf{y}_{1,g}^\prime \odot \dot{\mathbf{y}}_{2,g}^\prime - \mathbf{y}_{2,g}^\prime \odot \dot{\mathbf{y}}_{1,g}^\prime\right) \right)\right)
            \end{align}
            where $\boldsymbol{\alpha}_g$ is the angle between the global and local north vector that is defined with respect to a rotation around the $\mathbf{y}_g^\prime$ axis so that
            \begin{align}
                \boldsymbol{\alpha}_g = -\arcsin\left(\left(\mathbf{x}_{2,g}^\prime d_3 - \mathbf{x}_{3,g}^\prime d_2\right) \odot \mathbf{y}_{1,g}^\prime + \left(\mathbf{x}_{3,g}^\prime d_1 - \mathbf{x}_{1,g}^\prime d_3\right) \odot \mathbf{y}_{2,g}^\prime + \left(\mathbf{x}_{1,g}^\prime d_2 - \mathbf{x}_{2,g}^\prime d_1\right) \odot \mathbf{y}_{3,g}^\prime\right)
            \end{align}
            and $\sin\left(\boldsymbol{\alpha}_g\right)$ denotes an element wise function evaluation. The derivation of the angular velocity $\dot{\boldsymbol{\varphi}}_g^\prime$ around the local, magnetic north vector is subsequently outlined. The quaternion for the rotation from the drone fixed coordinate system to the local coordinate system that is oriented to the magnetic north vector is
            \begin{align}
                \mathbf{q}_g =
                \left[
                \begin{array}{cccc}
                    \cos\left(\frac{1}{2} \boldsymbol{\alpha}_g\right) &
                    \sin\left(\frac{1}{2} \boldsymbol{\alpha}_g\right) \odot {\mathbf{y}_{1,g}^\prime}^\top &
                    \sin\left(\frac{1}{2} \boldsymbol{\alpha}_g\right) \odot {\mathbf{y}_{2,g}^\prime}^\top &
                    \sin\left(\frac{1}{2} \boldsymbol{\alpha}_g\right) \odot {\mathbf{y}_{3,g}^\prime}^\top
                \end{array}
                \right]^\top.
            \end{align}
            The angular velocity from this transformation for the $i$-th element is
            \begin{align}
                \left[
                \begin{array}{c}
                    0\\
                    \dot{\boldsymbol{\varphi}}_g^{\star\ i}
                \end{array}
                \right] = 2 \dot{\mathbf{q}}_g^i \tilde{\mathbf{q}}_g^i
            \end{align}
            where $\tilde{\mathbf{q}}_g^i$ is the conjugated quaternion. The angular velocity $\dot{\boldsymbol{\varphi}}^\prime_g$ can therefore be written as
            \begin{align}
                \dot{\boldsymbol{\varphi}}^\prime_g =
                -\mathbf{x}_{1,g}^\prime \odot \left(\dot{\boldsymbol{\varphi}}_{x,g} + \dot{\boldsymbol{\varphi}}_{x,g}^\star \right)
                -\mathbf{x}_{2,g}^\prime \odot \left(\dot{\boldsymbol{\varphi}}_{y,g} + \dot{\boldsymbol{\varphi}}_{y,g}^\star \right)
                -\mathbf{x}_{3,g}^\prime \odot \left(\dot{\boldsymbol{\varphi}}_{z,g} + \dot{\boldsymbol{\varphi}}_{z,g}^\star \right).
            \end{align}


    \section{Global Orientation and Position}
        \subsection{Sensor Data}
            The data streams for the third and final stage are subsequently summarized. The data stream of the GNSS receiver is
            \begin{align}
                \left[
                \begin{array}{c}
                    \mathbf{t}_G\\
                    \mathbf{x}_G\\
                    \mathbf{y}_G\\
                    \mathbf{z}_G
                \end{array}
                \right] =
                \left[
                \begin{array}{ccccc}
                    \mathbf{t}_{G,1} & \mathbf{t}_{G,2} & \mathbf{t}_{G,3} & \cdots & \mathbf{t}_{G,n_G}\\
                    \mathbf{x}_{G,1} & \mathbf{x}_{G,2} & \mathbf{x}_{G,3} & \cdots & \mathbf{x}_{G,n_G}\\
                    \mathbf{y}_{G,1} & \mathbf{y}_{G,2} & \mathbf{y}_{G,3} & \cdots & \mathbf{y}_{G,n_G}\\
                    \mathbf{z}_{G,1} & \mathbf{z}_{G,2} & \mathbf{z}_{G,3} & \cdots & \mathbf{z}_{G,n_G}
                \end{array}
                \right]
            \end{align}
            and the data stream of the barometer is
            \begin{align}
                \left[
                \begin{array}{c}
                    \mathbf{t}_B\\
                    \mathbf{z}_B
                \end{array}
                \right] =
                \left[
                \begin{array}{ccccc}
                    \mathbf{t}_{B,1} & \mathbf{t}_{B,2} & \mathbf{t}_{B,3} & \cdots & \mathbf{t}_{B,n_B}\\
                    \mathbf{z}_{B,1} & \mathbf{z}_{B,2} & \mathbf{z}_{B,3} & \cdots & \mathbf{z}_{B,n_B}
                \end{array}
                \right].
            \end{align}
            The transformed data streams of the accelerometers is
            \begin{align}
                \left[
                \begin{array}{c}
                    \mathbf{t}_a\\
                    \ddot{\mathbf{x}}_a^\prime\\
                    \ddot{\mathbf{y}}_a^\prime\\
                    \ddot{\mathbf{z}}_a^\prime
                \end{array}
                \right] =
                \left[
                \begin{array}{ccccc}
                    \mathbf{t}_{a,1} & \mathbf{t}_{a,2} & \mathbf{t}_{a,3} & \cdots & \mathbf{t}_{a,n_a}\\
                    \ddot{\mathbf{x}}_{a,1}^\prime & \mathbf{x}_{a,2}^\prime & \mathbf{x}_{a,3}^\prime & \cdots & \mathbf{x}_{a,n_a}^\prime\\
                    \ddot{\mathbf{y}}_{a,1}^\prime & \mathbf{y}_{a,2}^\prime & \mathbf{y}_{a,3}^\prime & \cdots & \mathbf{y}_{a,n_a}^\prime\\
                    \ddot{\mathbf{z}}_{a,1}^\prime & \mathbf{z}_{a,2}^\prime & \mathbf{z}_{a,3}^\prime & \cdots & \mathbf{z}_{a,n_a}^\prime
                \end{array}
                \right]
            \end{align}
            and the transformed angular velocity is
            \begin{align}
                \left[
                \begin{array}{c}
                    \mathbf{t}_g\\
                    \dot{\boldsymbol{\varphi}}_g^\prime
                \end{array}
                \right] =
                \left[
                \begin{array}{ccccc}
                    \mathbf{t}_{g,1} & \mathbf{t}_{g,2} & \mathbf{t}_{g,3} & \cdots & \mathbf{t}_{g,n_g}\\
                    \dot{\boldsymbol{\varphi}}_{g,1}^\prime & \dot{\boldsymbol{\varphi}}_{g,2}^\prime & \dot{\boldsymbol{\varphi}}_{g,3}^\prime & \cdots & \dot{\boldsymbol{\varphi}}_{g,n_g}^\prime
                \end{array}
                \right].
            \end{align}


        \subsection{Error Terms}
            A fusion of previously summarized data streams leads to an accurate position and orientation estimation. For example, the shape functions for the the coordinates, accelerations and the rotation angle are subsequently denoted as
            \begin{align}
                \mathbf{x}_G^\prime = \mathbf{N}_G \mathbf{R}_x \mathbf{u} = \mathbf{N}_{G,x} \mathbf{u},\hspace{5mm}
                \dot{\boldsymbol{\varphi}}^{\prime\prime}_g = \dot{\mathbf{N}}_g \mathbf{R}_\varphi \mathbf{u} = \dot{\mathbf{N}}_{g,\varphi} \mathbf{u}\hspace{5mm}\textrm{and}\hspace{5mm}
                \ddot{\mathbf{x}}^{\prime\prime\prime}_a = \ddot{\mathbf{N}}_a \mathbf{R}_x \mathbf{u} = \ddot{\mathbf{N}}_{a,x} \mathbf{u}.
            \end{align}
            where it should be noted that $\mathbf{R}_x$ and $\mathbf{u}$ differ from the terms that are used for the computation of the local magnetic north vector during the first stage. Furthermore, the different number of primes is due to the varying number of transformations needed to compute the corresponding terms. The interpolation error $E$ of the sensor data is
            \begin{align}
                E = &\hspace{3.5mm} \frac{1}{2 \sigma_G^2}\left(
                \left(\mathbf{x}_G^\prime - \mathbf{x}_G\right)^\top \left(\mathbf{x}_G^\prime - \mathbf{x}_G\right) + \left(\mathbf{y}_G^\prime - \mathbf{y}_G\right)^\top \left(\mathbf{y}_G^\prime - \mathbf{y}_G\right) + \left(\mathbf{z}_G^\prime - \mathbf{z}_G\right)^\top \left(\mathbf{z}_G^\prime - \mathbf{z}_G\right)\right)\\\nonumber
                &+ \frac{1}{2 \sigma_B^2}\left(\mathbf{z}_B^\prime - \mathbf{z}_B\right)^\top \left(\mathbf{z}_B^\prime - \mathbf{z}_B\right) + \frac{1}{2 \overline{\sigma}_g^2}\left(\dot{\boldsymbol{\varphi}}^{\prime\prime}_g - \dot{\boldsymbol{\varphi}}^\prime_g\right)^\top \left(\dot{\boldsymbol{\varphi}}^{\prime\prime}_g - \dot{\boldsymbol{\varphi}}^\prime_g\right)\\\nonumber
                &+ \frac{1}{2 \overline{\sigma}_a^2}\left(\left(\ddot{\mathbf{x}}_a^{\prime\prime\prime} - \ddot{\mathbf{x}}_a^{\prime\prime}\right)^\top \left(\ddot{\mathbf{x}}_a^{\prime\prime\prime} - \ddot{\mathbf{x}}_a^{\prime\prime}\right) + \left(\ddot{\mathbf{y}}_a^{\prime\prime\prime} - \ddot{\mathbf{y}}_a^{\prime\prime}\right)^\top \left(\ddot{\mathbf{y}}_a^{\prime\prime\prime} - \ddot{\mathbf{y}}_a^{\prime\prime}\right) + \left(\ddot{\mathbf{z}}_a^{\prime\prime\prime} - \ddot{\mathbf{z}}_a^{\prime\prime}\right)^\top \left(\ddot{\mathbf{z}}_a^{\prime\prime\prime} - \ddot{\mathbf{z}}_a^{\prime\prime}\right)\right)
            \end{align}
            where the only nonlinear terms are due to the accelerations since
            \begin{align}
                \left[
                \begin{array}{c}
                    \ddot{\mathbf{x}}_a^{\prime\prime}\\
                    \ddot{\mathbf{y}}_a^{\prime\prime}\\
                    \ddot{\mathbf{z}}_a^{\prime\prime}
                \end{array}
                \right] =
                \left[
                \begin{array}{c}
                    \mathbf{X}_{1,a} \odot \ddot{\mathbf{x}}_a^\prime + \left(\mathbf{Y}_{1,a} \odot \ddot{\mathbf{y}}_a^\prime + \mathbf{Z}_{1,a} \odot \ddot{\mathbf{z}}_{a}^\prime\right) \odot \cos\left(\boldsymbol{\varphi}^{\prime\prime}_a\right) + \left(\mathbf{Y}_{1,a} \odot \ddot{\mathbf{z}}_a^\prime - \mathbf{Z}_{1,a} \odot \ddot{\mathbf{y}}_{a}^\prime\right) \odot \sin\left(\boldsymbol{\varphi}^{\prime\prime}_a\right)\\
                    \mathbf{X}_{2,a} \odot \ddot{\mathbf{x}}_a^\prime + \left(\mathbf{Y}_{2,a} \odot \ddot{\mathbf{y}}_a^\prime + \mathbf{Z}_{2,a} \odot \ddot{\mathbf{z}}_{a}^\prime\right) \odot \cos\left(\boldsymbol{\varphi}^{\prime\prime}_a\right) + \left(\mathbf{Y}_{2,a} \odot \ddot{\mathbf{z}}_a^\prime - \mathbf{Z}_{2,a} \odot \ddot{\mathbf{y}}_{a}^\prime\right) \odot \sin\left(\boldsymbol{\varphi}^{\prime\prime}_a\right)\\
                    \mathbf{X}_{3,a} \odot \ddot{\mathbf{x}}_a^\prime + \left(\mathbf{Y}_{3,a} \odot \ddot{\mathbf{y}}_a^\prime + \mathbf{Z}_{3,a} \odot \ddot{\mathbf{z}}_{a}^\prime\right) \odot \cos\left(\boldsymbol{\varphi}^{\prime\prime}_a\right) + \left(\mathbf{Y}_{3,a} \odot \ddot{\mathbf{z}}_a^\prime - \mathbf{Z}_{3,a} \odot \ddot{\mathbf{y}}_{a}^\prime\right) \odot \sin\left(\boldsymbol{\varphi}^{\prime\prime}_a\right)
                \end{array}
                \right].
            \end{align}
            The first derivative of the error term is
            \begin{align}
                \frac{\partial E}{\partial \mathbf{u}} = &\hspace{3.5mm} \frac{1}{\sigma_G^2} \left( {\mathbf{N}_{G,x}}^\top \left(\mathbf{N}_{G,x} \mathbf{u} - \mathbf{x}_G\right) + {\mathbf{N}_{G,y}}^\top \left(\mathbf{N}_{G,y} \mathbf{u} - \mathbf{y}_G\right) + {\mathbf{N}_{G,z}}^\top \left(\mathbf{N}_{G,z} \mathbf{u} - \mathbf{z}_G\right) \right)\\\nonumber
                &+ \frac{1}{\sigma_B^2} {\mathbf{N}_{B,z}}^\top\left(\mathbf{N}_{B,z} \mathbf{u} - \mathbf{z}_B\right) + \frac{1}{\overline{\sigma}_g^2} {\dot{\mathbf{N}}_{g,\varphi}}^\top\left(\dot{\mathbf{N}}_{g,\varphi} \mathbf{u} - \dot{\boldsymbol{\varphi}}_g^\prime\right)\\\nonumber
                &+ \frac{1}{\overline{\sigma}_a^2} \left(\left(\ddot{\mathbf{N}}_{a,x} - \frac{\partial \ddot{\mathbf{x}}_a^{\prime\prime}}{\partial \mathbf{u}}\right)^\top \left(\ddot{\mathbf{N}}_{a,x} \mathbf{u} - \ddot{\mathbf{x}}_a^{\prime\prime}\right) + \left(\ddot{\mathbf{N}}_{a,y} - \frac{\partial \ddot{\mathbf{y}}_a^{\prime\prime}}{\partial \mathbf{u}}\right)^\top \left(\ddot{\mathbf{N}}_{a,y} \mathbf{u} - \ddot{\mathbf{y}}_a^{\prime\prime}\right) + \left(\ddot{\mathbf{N}}_{a,z} - \frac{\partial \ddot{\mathbf{z}}_a^{\prime\prime}}{\partial \mathbf{u}}\right)^\top \left(\ddot{\mathbf{N}}_{a,z} \mathbf{u} - \ddot{\mathbf{z}}_a^{\prime\prime}\right)\right)
            \end{align}
            where, for example
            \begin{align}
                \frac{\partial \ddot{\mathbf{x}}_a^{\prime\prime}}{\partial \mathbf{u}} =
                \textrm{diag}\left(\left(\mathbf{Y}_{1,a} \odot \ddot{\mathbf{z}}_a^\prime - \mathbf{Z}_{1,a} \odot \ddot{\mathbf{y}}_{a}^\prime\right) \odot \cos\left(\boldsymbol{\varphi}^{\prime\prime}_a\right)
                - \left(\mathbf{Y}_{1,a} \odot \ddot{\mathbf{y}}_a^\prime + \mathbf{Z}_{1,a} \odot \ddot{\mathbf{z}}_{a}^\prime\right) \odot \sin\left(\boldsymbol{\varphi}^{\prime\prime}_a\right)\right) \mathbf{N}_{a,\varphi}.
            \end{align}
            The corresponding second order derivative is
            \begin{align}
                \frac{\partial^2 E}{\partial \mathbf{u}^2} = &\hspace{2mm} \frac{1}{\sigma_G^2} \left( {\mathbf{N}_{G,x}}^\top \mathbf{N}_{G,x} + {\mathbf{N}_{G,y}}^\top \mathbf{N}_{G,y} + {\mathbf{N}_{G,z}}^\top \mathbf{N}_{G,z} \right)
                + \frac{1}{\sigma_B^2} {\mathbf{N}_{B,z}}^\top \mathbf{N}_{B,z} + \frac{1}{\overline{\sigma}_g^2} {\dot{\mathbf{N}}_{g,\varphi}}^\top \dot{\mathbf{N}}_{g,\varphi}\\\nonumber
                &\hspace{-1.5mm} + \frac{1}{\overline{\sigma}_a^2} \left(\hspace{3.5mm} \left(\ddot{\mathbf{N}}_{a,x} - \frac{\partial \ddot{\mathbf{x}}_a^{\prime\prime}}{\partial \mathbf{u}}\right)^\top \left(\ddot{\mathbf{N}}_{a,x} - \frac{\partial \ddot{\mathbf{x}}_a^{\prime\prime}}{\partial \mathbf{u}} \right) +
                \left(\ddot{\mathbf{N}}_{a,y} - \frac{\partial \ddot{\mathbf{y}}_a^{\prime\prime}}{\partial \mathbf{u}}\right)^\top \left(\ddot{\mathbf{N}}_{a,y} - \frac{\partial \ddot{\mathbf{y}}_a^{\prime\prime}}{\partial \mathbf{u}}\right) + \left(\ddot{\mathbf{N}}_{a,z} - \frac{\partial \ddot{\mathbf{z}}_a^{\prime\prime}}{\partial \mathbf{u}}\right)^\top \left(\ddot{\mathbf{N}}_{a,z} - \frac{\partial \ddot{\mathbf{z}}_a^{\prime\prime}}{\partial \mathbf{u}}\right)\right.\\\nonumber
                &\hspace{7.5mm}
                + {\mathbf{N}_{a,\varphi}}^\top \textrm{diag}\left(\hspace{2mm}
                \left(\ddot{\mathbf{N}}_{a,x} \mathbf{u} - \ddot{\mathbf{x}}_a^{\prime\prime}\right) \odot \left( \left(\mathbf{Y}_{1,a} \odot \ddot{\mathbf{y}}_a^{\prime\prime} + \mathbf{Z}_{1,a} \odot \ddot{\mathbf{z}}_a^{\prime\prime}\right) \odot \cos\left(\boldsymbol{\varphi}_a^{\prime\prime}\right) + \left(\mathbf{Y}_{1,a} \odot \ddot{\mathbf{z}}^{\prime\prime}_a - \mathbf{Z}_{1,a} \odot \ddot{\mathbf{y}}_a^{\prime\prime}\right) \odot \sin\left(\boldsymbol{\varphi}_a^{\prime\prime}\right) \right) \right.\\\nonumber
                & \hspace{25mm} + \left(\ddot{\mathbf{N}}_{a,y} \mathbf{u} - \ddot{\mathbf{y}}_a^{\prime\prime}\right) \odot \left( \left(\mathbf{Y}_{2,a} \odot \ddot{\mathbf{y}}_a^{\prime\prime} + \mathbf{Z}_{2,a} \odot \ddot{\mathbf{z}}_a^{\prime\prime}\right) \odot \cos\left(\boldsymbol{\varphi}_a^{\prime\prime}\right) + \left(\mathbf{Y}_{2,a} \odot \ddot{\mathbf{z}}^{\prime\prime}_a - \mathbf{Z}_{2,a} \odot \ddot{\mathbf{y}}_a^{\prime\prime}\right) \odot \sin\left(\boldsymbol{\varphi}_a^{\prime\prime}\right) \right)\\\nonumber
                & \left.\hspace{25mm} + \left(\ddot{\mathbf{N}}_{a,z} \mathbf{u} - \ddot{\mathbf{z}}_a^{\prime\prime}\right) \odot \left( \left(\mathbf{Y}_{3,a} \odot \ddot{\mathbf{y}}_a^{\prime\prime} + \mathbf{Z}_{3,a} \odot \ddot{\mathbf{z}}_a^{\prime\prime}\right) \odot \cos\left(\boldsymbol{\varphi}_a^{\prime\prime}\right) + \left(\mathbf{Y}_{3,a} \odot \ddot{\mathbf{z}}^{\prime\prime}_a - \mathbf{Z}_{3,a} \odot \ddot{\mathbf{y}}_a^{\prime\prime}\right) \odot \sin\left(\boldsymbol{\varphi}_a^{\prime\prime}\right) \right)
                \right) \mathbf{N}_{a,\varphi}.
            \end{align}
            The solution approach for the computation of the global orientation and position is identical to the approach that is used for the first stage.


    \section{Example}
        \subsection{Orientation}
            It is subsequently assumed that the time dependent orientation of the drone is described by the rotation vector $\mathbf{v}$
            \begin{align}
                \mathbf{v} =
                \left[
                \begin{array}{ccc}
                    0.7 & 0 & 0\\
                    0 & 0.3 & 0\\
                    0 & 0 & 0.9
                \end{array}
                \right]
                \left[
                \begin{array}{l}
                    \sin\left(\pi/4 + \hspace{2.5mm} 2 \pi t\right)\\
                    \sin\left(\pi/2 + 0.8 \pi t\right)\\
                    \sin\left(\hspace{11.5mm} 2 \pi t\right)
                \end{array}
                \right]
            \end{align}
            and a rotation angle $\varphi$
            \begin{align}
                \varphi = \frac{\pi}{2} \sin\left(\frac{\pi}{4} + 2.4 \pi t\right)
            \end{align}
            so that the corresponding quaternion is
            \begin{align}
                \mathbf{q} =
                \left[
                \begin{array}{l}
                    \cos\left(\varphi/2\right)\\
                    \sin\left(\varphi/2\right) \mathbf{v}/\|\mathbf{v}\|
                \end{array}
                \right].
            \end{align}


        \subsection{Position}
            Similarly, it is assumed that the time dependent sensor position is described by
            \begin{align}
                \left[
                \begin{array}{c}
                    X\\
                    Y\\
                    Z
                \end{array}
                \right] =
                \left[
                \begin{array}{c}
                    0.1\\ 0\\ 0
                \end{array}
                \right] t +
                \left[
                \begin{array}{ccc}
                    0.3 & 0 & 0\\
                    0 & 1.0 & 0\\
                    0 & 0 & 1.3
                \end{array}
                \right]
                \left[
                \begin{array}{l}
                    \sin\left(\pi/2 + 2 \pi t\right)\\
                    \sin\left(\hspace{9mm} 2 \pi t\right)\\
                    \sin\left(\pi/4 + \hspace{2mm} \pi t\right)
                \end{array}
                \right].
            \end{align}


        \subsection{Sensor Data}
            The sensor data is derived from the previously introduced position and orientation terms. The standard deviations and sampling frequencies of the considered sensors are
            \begin{center}
                \begin{tabular}{c|cc}
                    Sensor & Standard Deviation ($\sigma$) & Sampling Frequency ($f$)\\\hline
                    GNSS & 1.0~m & 30~Hz\\
                    Barometer & 0.1~m & 100~Hz\\
                    Magnetometer & 0.1 & 100~Hz\\
                    Accelerometer & 0.01~m/s$^2$ & 1~kHz\\
                    Gyroscope & 0.01~rad/s & 1~kHz.
                \end{tabular}
            \end{center}
            Furthermore, the standard deviation of the sampling times is assumed to be $\sigma_t = 10^{-4}$~s and the timespan of a single finite element is 0.05~s so that the update frequency is 20~Hz. It should be noted that a relatively high GNSS sampling frequency is assumed since the first element needs at least two position measurements in order to avoid a singular matrix. This is not a problem for subsequent elements since they are augmented with the condensed error matrix of the previous elements. Hence it would be possible to use relatively low GNSS sampling frequencies if the timespan of the first element is sufficiently large. The shape functions that are considered for the interpolation of the coordinates are based on six Hermitian polynomials whereas the shape functions for the rotation angle and vector are based on four Hermitian polynomials.\\

            The assumed global north vector $\mathbf{d}$ (that needs to be normalized) is
            \begin{align}
                \mathbf{d} =
                \left[
                \begin{array}{ccc}
                    1 & 0.1 & 0.2
                \end{array}
                \right]^\top
            \end{align}
            and the assumed standard deviations for the first stage are
            \begin{align}
                \sigma_L = 10^{-1.5}~\sigma_g,\hspace{5mm}
                \overline{\sigma}_g = 10^{1.5} \sigma_g
            \end{align}
            and for the third stage are
            \begin{align}
                \overline{\sigma}_g = 10^1 \sigma_g,\hspace{5mm}
                \overline{\sigma}_a = 10^2 \sigma_a.
            \end{align}


        \subsection{First Stage}
\begin{figure}[htbp]
                \begin{center}
                    \subfloat[]{
                        \includegraphics[width=\textwidth]{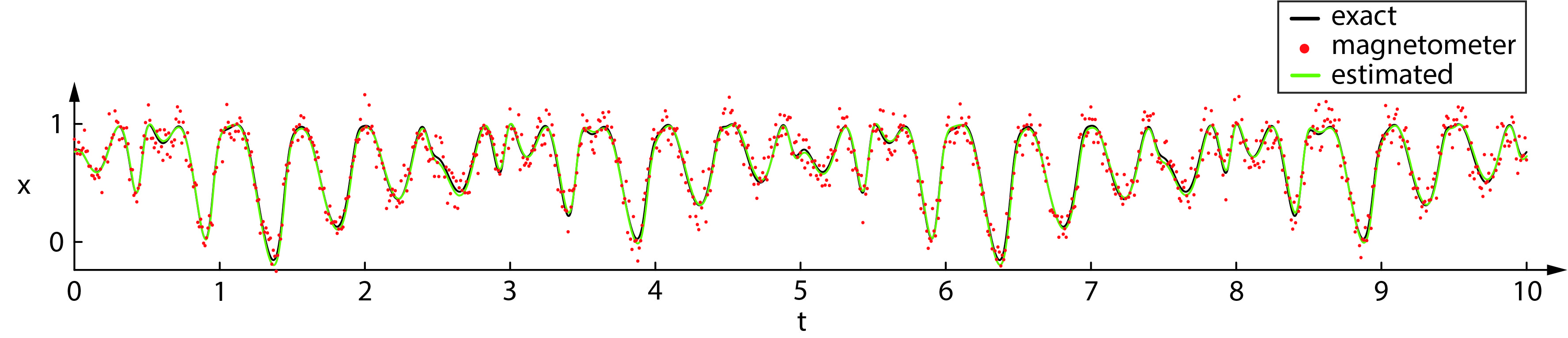}}

                    \subfloat[]{
                        \includegraphics[width=\textwidth]{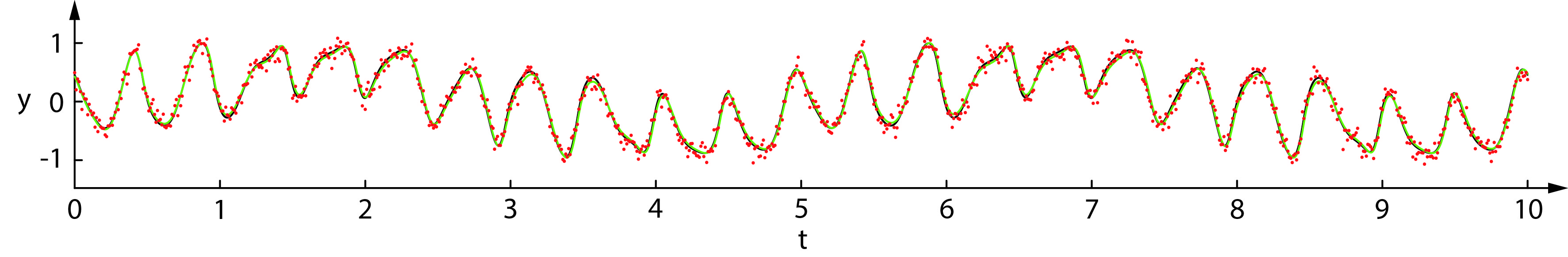}}

                    \subfloat[]{
                        \includegraphics[width=\textwidth]{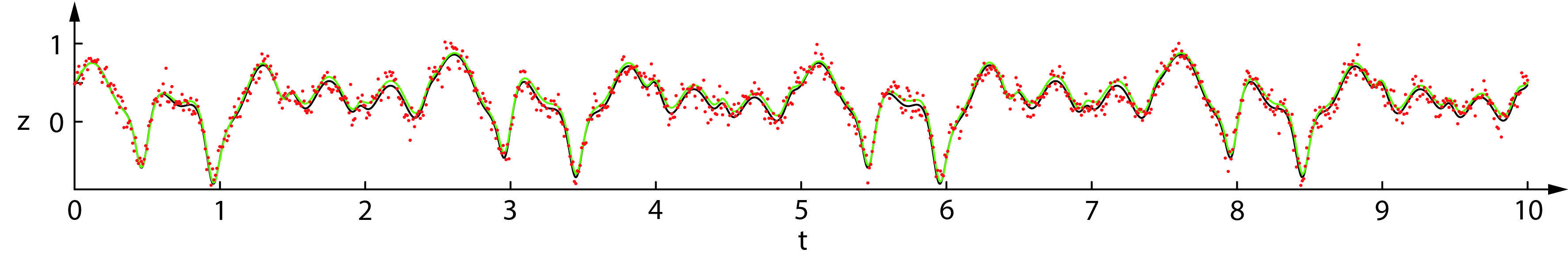}}

                    \subfloat[]{
                        \includegraphics[width=\textwidth]{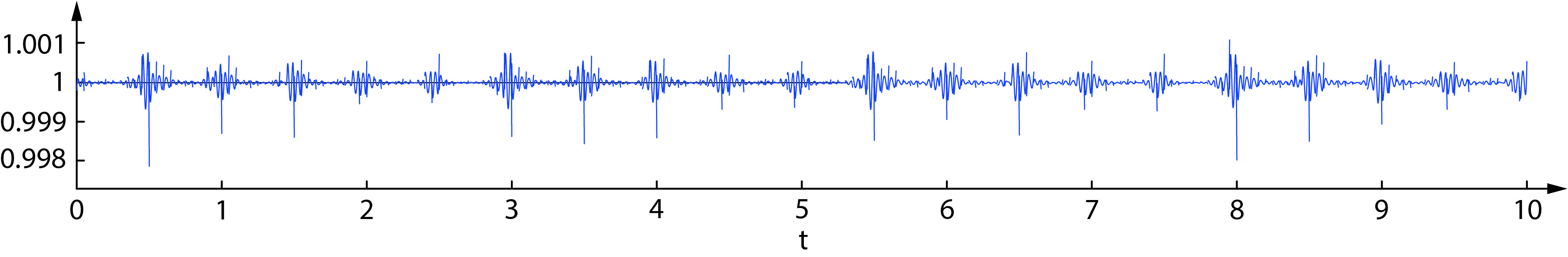}}
                    \caption{Estimated magnetic north vector of the first stage. Vector components (a) x, (b) y and (c) z. (d) Vector length.}
                  	\label{pic:Figure_1}
                \end{center}
            \end{figure}

            The magnetic north vector that is computed in the first stage is illustrated in Figure~\ref{pic:Figure_1}. It can be seen that the estimation closely reassembles the exact solution despite relatively noisy magnetometer measurements. Furthermore, the penalty based length constraint leads to a nearly unit vector length.


        \subsection{Third Stage}
\begin{figure}[htbp]
                \begin{center}
                    \subfloat[]{
                        \includegraphics[width=\textwidth]{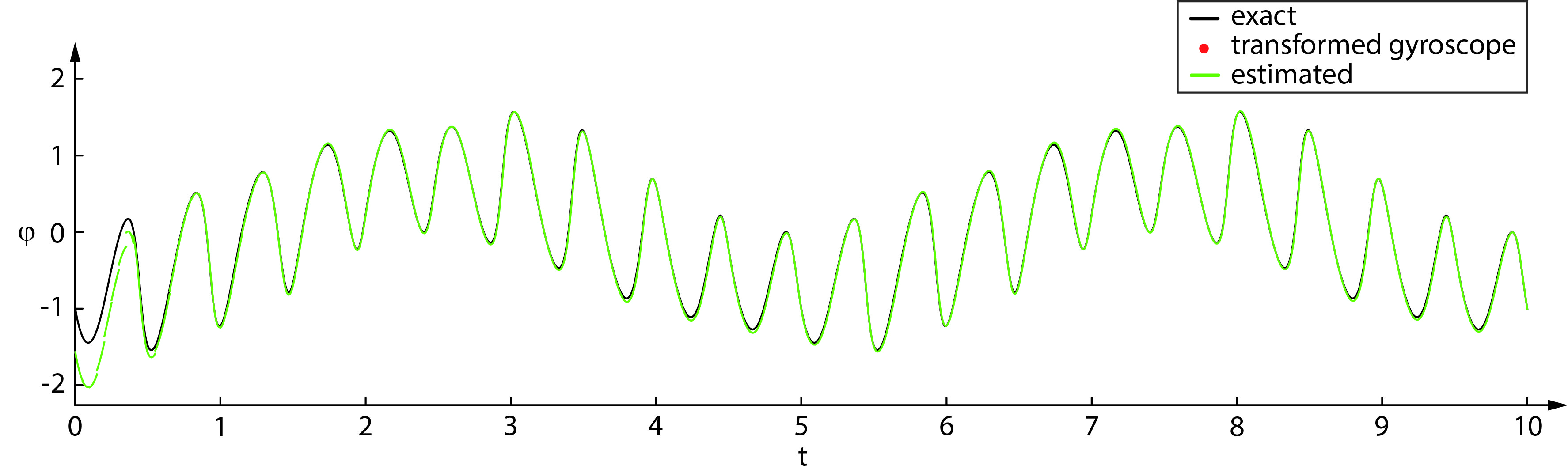}}

                    \subfloat[]{
                        \includegraphics[width=\textwidth]{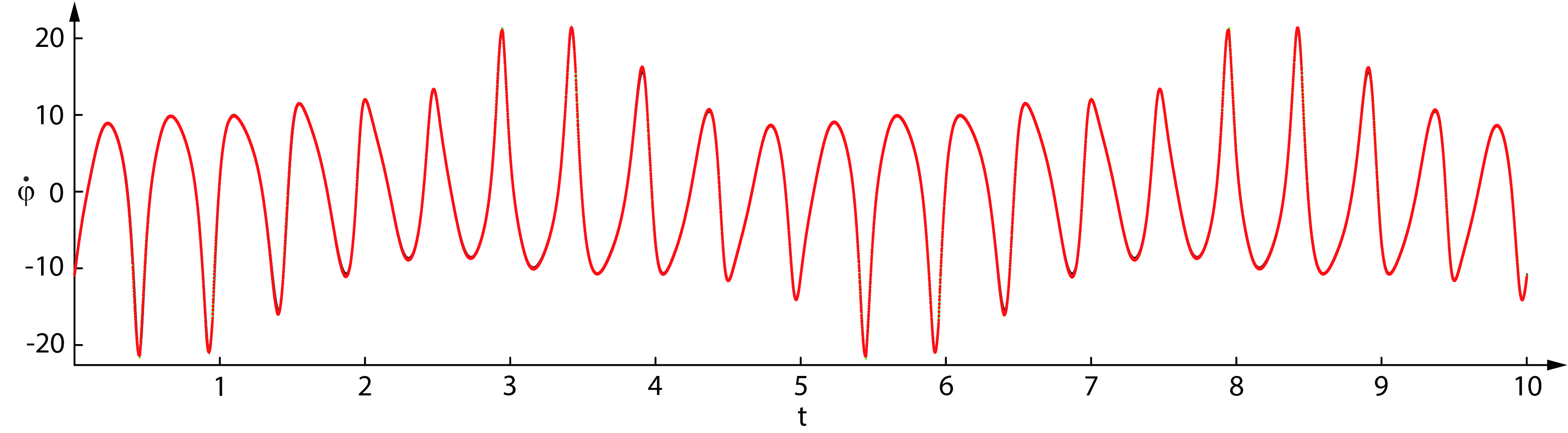}}
                    \caption{(a) Rotation angle and (b) rotational velocity around magnetic north vector.}
                  	\label{pic:Figure_2}
                \end{center}
            \end{figure}

\begin{figure}[htbp]
                \begin{center}
                    \subfloat[]{
                        \includegraphics[width=\textwidth]{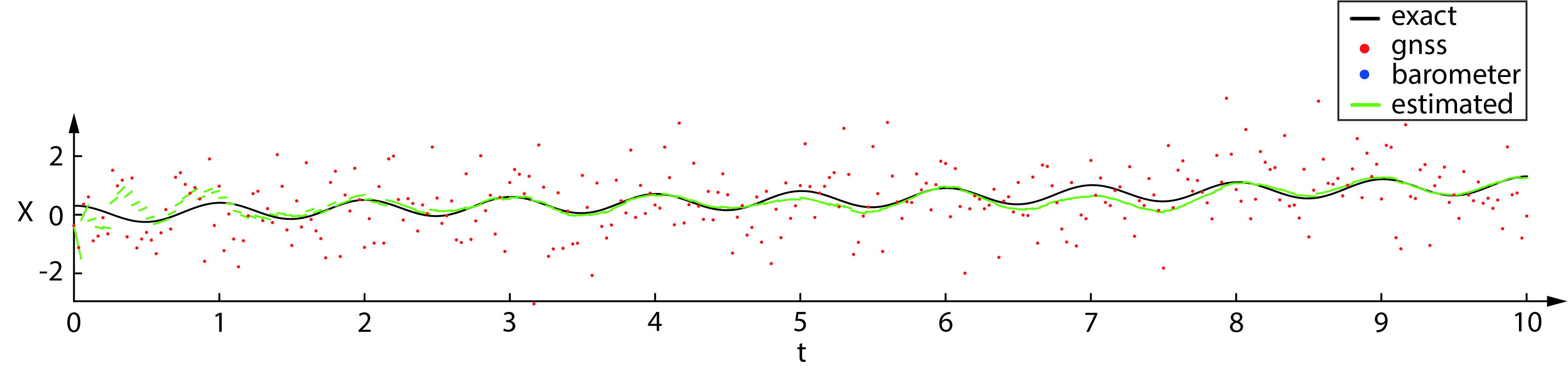}}

                    \subfloat[]{
                        \includegraphics[width=\textwidth]{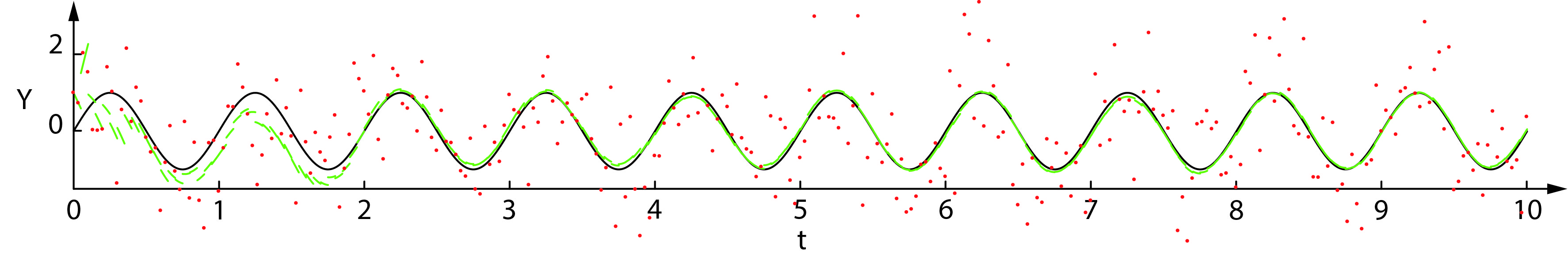}}

                    \subfloat[]{
                        \includegraphics[width=\textwidth]{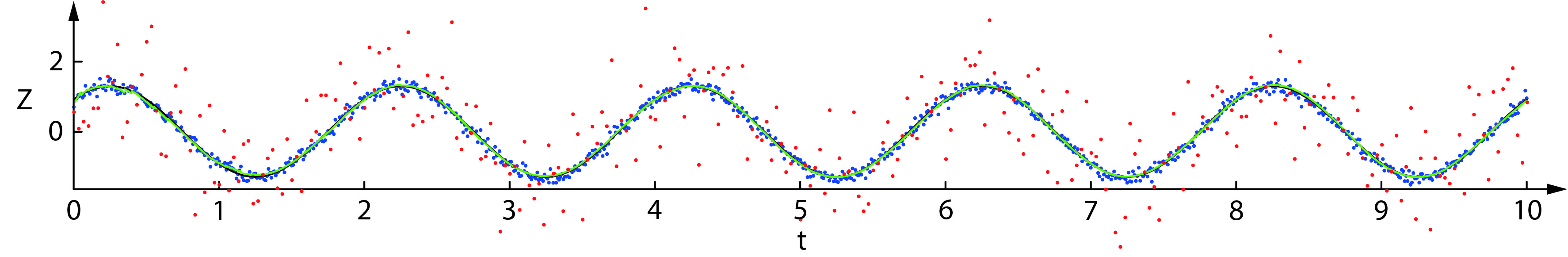}}
                    \caption{Estimated position. (a) X, (b) Y and (c) Z coordinates.}
                  	\label{pic:Figure_3}
                \end{center}
            \end{figure}

            The results of the third stage are shown in Figure~\ref{pic:Figure_2} and Figure~\ref{pic:Figure_3}. It can be seen that the initial estimates of the rotation angle and coordinates differ considerably from the exact solution. This is due to the large GNSS noise and the very limited amount of data available for the first few elements. Nonetheless, the estimated rotation angle and position converges to the exact solution after about two seconds.

        \subsection{Runtime}
            The presented algorithm was implemented in Matlab and executed on a single Intel i5-4250U CPU core. The processor utilization for a 20~Hz update rate in realtime was less than 25~\%. Hence it seems possible to achieve update rates of more than 100~Hz since higher update rates decrease the required number of iterations for each element.


    \section{Conclusion}
        The proposed method uses a space time finite element to nonlinearly fuse the data of a GNSS receiver, barometer, magnetometers, accelerometers and gyroscopes. It was shown that the influence, gain of previous sensor readings on the state of the current finite element can be elegantly taken into account with the help of static condensation. Furthermore, it was shown that a Matlab implementation of this algorithm can achieve realtime update rates of more than 100~Hz on relatively limited hardware.


    \appendix
    \section{Shape Functions}
        The shape functions are based on Hermitian polynomials that are composed from Lagrange polynomials. The latter can be written as
        \begin{align}
            L_j\left(\xi\right) = \prod_{\substack{0\leq m\leq k \\ m \neq j}} \frac{\xi-\xi_m}{\xi_j-\xi_m}
        \end{align}
        where $\xi\in\left[-1,1\right]$ and $\xi_j$, $\xi_m$ are the abscissa of the equally spaced roots. The corresponding first and second derivatives are
        \begin{align}
            \dot{L}_j\left(\xi\right) = \sum_{\substack{i=0 \\ i\neq j}}^k \left[ \frac{1}{x_j-x_i} \prod_{\substack{m=0\\ m\neq \left(i,j\right)}}^k \frac{x-x_m}{x_j-x_m} \right] \dot{\xi}
        \end{align}
        and
        \begin{align}
            \ddot{L}_j\left(\xi\right) = \sum_{\substack{i=0\\ i\neq j}}^k \frac{1}{x_j-x_i} \left[ \sum_{\substack{m=0\\ m\neq \left(i,j\right)}}^k \left(\frac{1}{x_j-x_m} \prod_{\substack{l=0\\ l\neq\left(i,j,m\right)}}^k \frac{x-x_l}{x_j-x_l}\right) \right] \dot{\xi}^2.
        \end{align}
        The Hermitian polynomials result in
        \begin{align}
            \mathbf{N}_k = \left(\mathbf{T}_k\right)^{-1} \mathbf{L}_k, \hspace{5mm} \mathbf{\dot{N}}_k = \left(\mathbf{T}_k\right)^{-1} \mathbf{\dot{L}}_k \hspace{5mm} \textrm{and} \hspace{5mm} \mathbf{\ddot{N}}_k = \left(\mathbf{T}_k\right)^{-1} \mathbf{\ddot{L}}_k
        \end{align}
        where the transformation matrix $\mathbf{T}$ for $k=3$ is
        \begin{align}
            \mathbf{T}_3 =
            \left[
            \begin{array}{cccc}
                L_0\left(-1\right) & L_1\left(-1\right) & L_2\left(-1\right) & L_3\left(-1\right)\\
                \dot{L}_0\left(-1\right) & \dot{L}_1\left(-1\right) & \dot{L}_2\left(-1\right) & \dot{L}_3\left(-1\right)\\\hline
                L_0\left(+1\right) & L_1\left(+1\right) & L_2\left(+1\right) & L_3\left(+1\right)\\
                \dot{L}_0\left(+1\right) & \dot{L}_1\left(+1\right) & \dot{L}_2\left(+1\right) & \dot{L}_3\left(+1\right)
            \end{array}
            \right]
        \end{align}
        and for $k=5$ is
        \begin{align}
            \mathbf{T}_5 =
            \left[
            \begin{array}{cccccc}
                L_0\left(-1\right) & L_1\left(-1\right) & L_2\left(-1\right) & L_3\left(-1\right) & L_4\left(-1\right) & L_5\left(-1\right)\\
                \dot{L}_0\left(-1\right) & \dot{L}_1\left(-1\right) & \dot{L}_2\left(-1\right) & \dot{L}_3\left(-1\right) & \dot{L}_4\left(-1\right) & \dot{L}_5\left(-1\right)\\
                \ddot{L}_0\left(-1\right) & \ddot{L}_1\left(-1\right) & \ddot{L}_2\left(-1\right) & \ddot{L}_3\left(-1\right) & \ddot{L}_4\left(-1\right) & \ddot{L}_5\left(-1\right)\\\hline
                L_0\left(+1\right) & L_1\left(+1\right) & L_2\left(+1\right) & L_3\left(+1\right) & L_4\left(+1\right) & L_5\left(+1\right)\\
                \dot{L}_0\left(+1\right) & \dot{L}_1\left(+1\right) & \dot{L}_2\left(+1\right) & \dot{L}_3\left(+1\right) & \dot{L}_4\left(+1\right) & \dot{L}_5\left(+1\right)\\
                \ddot{L}_0\left(+1\right) & \ddot{L}_1\left(+1\right) & \ddot{L}_2\left(+1\right) & \ddot{L}_3\left(+1\right) & \ddot{L}_4\left(+1\right) & \ddot{L}_5\left(+1\right)
            \end{array}
            \right].
        \end{align}

\end{document}